\documentclass[aps,prb,reprint]{revtex4-1}
\usepackage{amsmath}
\usepackage{graphicx}
\usepackage{units}  
\usepackage{xspace}
\usepackage{subfigure}
\usepackage{hyperref}

\newcommand{\ve}{\varepsilon}
\newcommand{\mb}{\mathbf}
\newcommand{\tb}{\textbf}
\newcommand{\beq}{\begin{equation}}
\newcommand{\eeq}{\end{equation}}
\newcommand{\bea}{\begin{eqnarray}}
\newcommand{\eea}{\end{eqnarray}}

\usepackage[usenames]{color}

\begin{document}

\bibliographystyle{apsrev}
 
\title{Anomalies in a slightly doped insulator with strong particle-hole asymmetry and narrow gap---the case for SmB$_6$?}

\author{Hridis K. Pal}
   \email{hridis.pal@iitb.ac.in}
   \altaffiliation[Present address: ] {Department of Physics, Indian Institute of Technology Bombay, Powai, Mumbai 400076, India.}  
\affiliation{LPS, CNRS UMR 8502, Universit\'{e} Paris-Sud, Universit\'{e} Paris-Saclay, 91405  Orsay Cedex, France}

\begin{abstract} 
SmB$_6$, known to be a Kondo insulator, has received intense scrutiny in recent years due to its paradoxical experimental signatures: while some quantities show an insulating behavior, others point to a metallic state. This has led to the conjecture that SmB$_6$ hosts nontrivial excitations within its bulk gap, and has spawned several theories to that effect. In principle, there exists an alternative possibility: the system is a metal but unusually with both metal- and insulator-like properties. Inspired by this possibility, I consider a minimal model of a Kondo insulator---a flat band hybridized with a parabolic band---that is slightly electron doped, i.e., the chemical potential is in the conduction band but close to the band edge. I show that, at the phenomenological level, the dc conductivity, ac conductivity, specific heat, and quantum oscillations within this model exhibit unusual behaviors that are, surprisingly, qualitatively consistent with those observed experimentally in SmB$_6$. The rapid change of band curvature around the chemical potential arising from the strong particle-hole asymmetry and the narrow gap in the model, a feature not usually encountered in the textbook cases of metals or insulators, is at the heart of the unusual behaviors. 
\end{abstract}

\maketitle 

\section{Introduction}

In spite of being studied for over half a century \cite{discovery1,discovery2}, several anomalous observations in the material SmB$_6$, known to be a Kondo insulator, have eluded a satisfactory understanding. The situation is paradoxical: whereas some observables behave as if the system is an insulator, others are consistent with it being a metal. A band gap is clearly visible in photoemission spectroscopy \cite{arpes1,arpes2} which manifests as an activated behavior in the dc resistivity---there is a steep increase in the resistivity with decreasing temperature. Nevertheless, at extremely low temperatures, the resistivity develops a plateau, signaling the onset of a metallic channel of conduction \cite{dc1,dc2}. Additionally, measurements of optical conductivity \cite{ac1}, specific heat \cite{spht0,spht1,spht2}, and quantum oscillations \cite{qosc1,qosc2}, among others, seem to point to the existence of a nonzero density of states within the gap.

The recent prediction of topological surface states in the bulk gap provided a possible way to end the deadlock \cite{dzero}. At temperatures less than the gap, the surface states provide a conduction channel, which could explain the appearance of the plateau in the resistivity. 
The success, however, was limited as it failed to explain other observations: careful analysis of data for the optical conductivity, specific heat, and quantum oscillations reveal that they can arise only from states that are 3D and of bulk origin. Additionally, these observables show features not typical of conventional metals which need to be accounted for as well: the optical conductivity decreases with frequency at high temperatures but increases with frequency at low temperatures \cite{ac1}; the specific heat is strikingly large, comparable to that of a heavy fermion system \cite{spht0,spht1}; and quantum oscillations show a temperature dependence that does not follow the Lifshitz-Kosevich theory valid for metals \cite{qosc2}.

\begin{figure}
\includegraphics[angle=0,width=0.80\columnwidth]{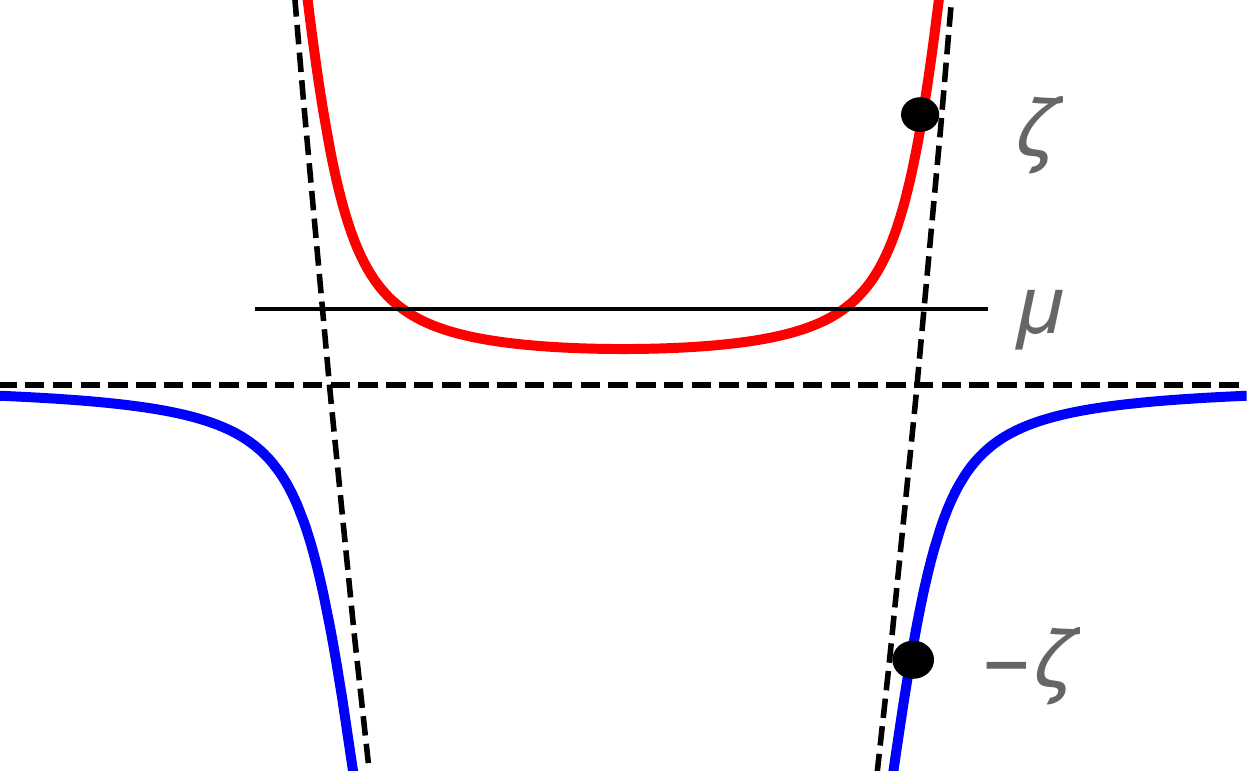}
\caption{Bands resulting from hybridization of a flat band with a parabolic band by a parameter $\zeta$, cf. Eq.~(\ref{ham}) (unhybridized bands are shown in dashes). The chemical potential is marked by $\mu$. Unlike in a Kondo insulator where it lies inside the gap, here it is assumed to lie in the conduction band close to the band edge. The rapid change of band curvature around $\mu$ results in unusual features in physical quantities as described in the text.
}
\label{band}
\end{figure}

In light of such findings, a consensus is slowly building up where the gap is thought to contain nontrivial excitations that are of bulk origin, in addition to the surface states. Several theories have been presented recently along this line of thinking \cite{baskaran,knolle2,erten2,chowdhury}. Regardless of the details, the starting point of all these theories is that the system, to begin with, is an insulator. In principle, however, there exists another possibility that has not received comparable attention: one can ask whether the system is instead a metal, where the conductivity somehow shows an insulator-like behavior with temperature. Of course, such a model will also have to explain the departures from the standard metallic behavior in the quantities mentioned above.

The goal of this paper is to explore whether the latter possibility is a reasonable one. To that effect I consider a minimal phenomenological model of a Kondo insulator---a parabolic band hybridized with a flat band---but assume that the chemical potential, instead of lying inside the gap as in a conventional Kondo insulator, lies inside the conduction band (valence band), close to the edge. I explore the phenomenology that results from such a model by calculating four quantities: dc conductivity, ac conductivity, specific heat, and quantum oscillations. Surprisingly, all of them feature anomalous behaviors that are qualitatively consistent with those observed in SmB$_6$. Their origin lies in the rapid change of band curvature around the chemical potential on the scale of temperature, a feature that is usually not encountered in textbook examples of metals or insulators. 

\section{Model}

Consider a model where a parabolic band hybridizes with a flat band due to a parameter $\zeta$. The Hamiltonian reads ($\hbar=k_B=1$)
\beq
H_{\mb{k}}=
\begin{pmatrix}
\frac{k^2}{2m}-\Delta&\zeta\\
\zeta&0
\end{pmatrix}.
\label{ham}
\eeq
Hybridization leads to avoided crossing which opens a gap. Here, $\Delta\gg\zeta>0$ determines the bandwidth of the valence band, and the gap is equal to $\zeta^2/\Delta\ll\zeta$---see Fig.~\ref{band}.When the chemical potential $\mu$ lies inside the gap, the system is an insulator.  Such a model provides a minimal description of a Kondo insulator. In the following, however, I consider a situation where $\mu$ is pushed slightly into the conduction band such that $\mu-E_c\ll\zeta$, where $E_c$ is the edge of the conduction band ($\mu$ in the valence band is also discussed for completeness). Such a choice for the model is inspired, in part, by a posteriori justification through the quantities calculated, and, in part, by experiment: photoemission spectroscopy presented in Ref.~\cite{support1} does find the chemical potential to be in the conduction band instead of the gap. Since this work focuses on the bulk bands, the momentum dependence in the hybridization term, important for topological properties of the surface states in the gap \cite{dzero}, is ignored. Also, effects of electronic interactions are not considered explicitly within this phenomenological model. It is assumed that the quasiparticles in this model are already renormalized due to interactions (see, however, Ref.~\cite{kats}). Nevertheless, interactions can give rise to further nontrivial effects that are not captured by this simple model. This requires a microscopic model and is outside the scope of this work.

\begin{figure}
\includegraphics[angle=0,width=0.95\columnwidth]{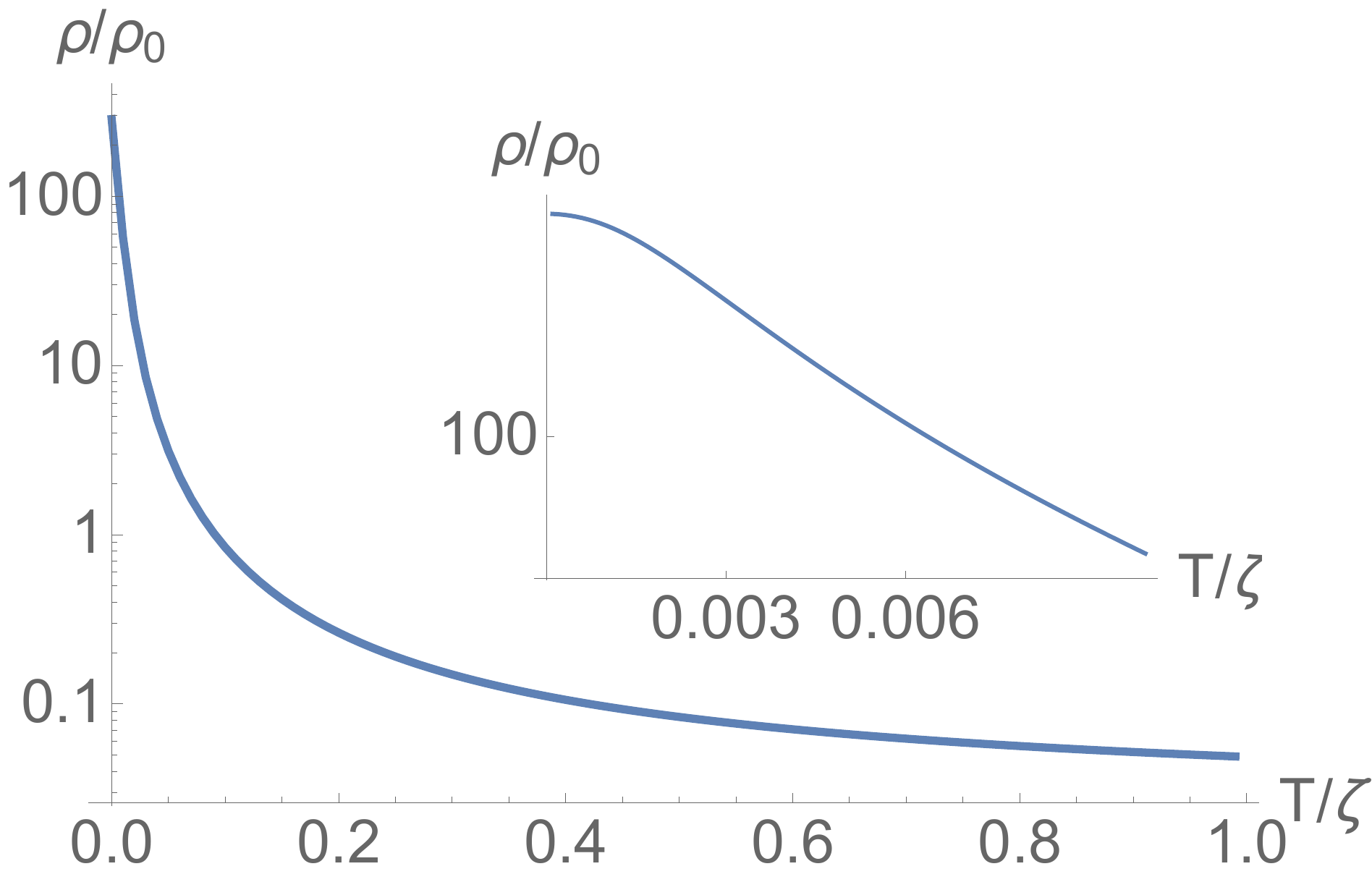}
\caption{Dependence of resistivity $\rho=1/\sigma$ on temperature $T$ according to Eq.~(\ref{dcconeq}) for $\zeta/\Delta=0.050$ and $\mu/\zeta=0.055$ ($\rho_0$: resistivity at $T=0$ for the unhybridized parabolic band). A steep increase with decreasing $T$ is observed that ends in a plateau at $T=0$ (shown in inset).}
\label{dctransport}
\end{figure}

\section{Physical Quantities}

Using the simple model above, I now calculate different physical quantities and demonstrate their anomalous behaviors.

\subsection{Dc conductivity } 

Within a constant relaxation time approximation, the Drude-Boltzmann conductivity can be calculated as 
\beq
\sigma=\sum_{i}\frac{2e^2\tau}{(2\pi)^3}\int v_i^2\left(-\frac{\partial f_0}{\partial\varepsilon_i}\right)d\mb{k},
\label{dcconeq}
\eeq
where $e$ is the electronic charge, $\tau$ is the scattering time, $v_i=\frac{\partial\ve_i}{\partial k}$, $f_0$ is the Fermi function, and the summation over $i$ runs over the two bands. Fig.~\ref{dctransport} presents the resistivity $\rho=1/\sigma$ as a function of temperature for the model in (\ref{ham}). With decrease in temperature in the regime $T\lesssim\zeta$, the resistivity increases sharply as in an insulator, but finally levels out into a plateau. The insulator-like behavior, in spite of the system being metallic, is not entirely unexpected. Since $\mu$ is very close to the edge of the band, the system is at the borderline between a metal and an insulator, and a truly metallic behavior is not expected. However, the mechanism by which this behavior arises here is distinct from that in conventional insulators. In the latter, increase in resistivity with decreasing temperature stems from a decrease in the number of carriers available. Here, with decreasing $T$, the region which contributes to the conductivity shrinks to the vicinity of the band edge. The density of states near the edge is high, i.e., there is no shortage of carriers; instead, the velocity of the carriers goes to zero leading to the increase in resistivity. However, when $T\ll\mu-E_c$, one enters a metallic regime which results in the plateau. In this picture, the temperature at which the plateau appears, $T_p$, and the corresponding resistivity, $\rho_p$, are related and arise from the scale $\mu-E_c$, but does not directly depend on $\zeta$. Note that the same behavior would arise if $\mu$ lies in the valence band.

It is instructive to compare with standard textbook results in terms of the Drude formula for conductivity, $\sigma=ne^2\tau/m$, where $n$ is the charge density and $m$ is the band mass \cite{ashcroft}. In standard metals, $n$ is large and does not change with $T$ which enters mainly through $\tau$. In insulators (semiconductors), it is the opposite: the dominant effect of $T$ enters through $n$. In semimetals, both $n$ and $\tau$ contribute to the $T-$dependence \cite{graphite}. The situation here does not belong to any of these categories. Here, the main $T-$dependence can be thought to arise effectively from $m$ changing with $T$.

A steep increase in resistivity that evolves into a plateau is a hallmark feature of SmB$_6$, reported in several experiments \cite{dc1,dc2}. While the increase is attributed to conventional insulating behavior due to the gap (with $\mu$ inside it), the plateau has been credited to midgap impurity states in the past, but more recently to topological surface states in the gap. The model presented here provides an alternative. Note, however, this does not necessarily imply that the surface states do not play a role. Close to the edge of the band where $\mu$ is assumed to lie, the surface bands merge with the bulk bands. It is possible that both channels contribute to conductivity, and because the bulk contribution is extremely small, the surface contribution could be comparable. This might explain why a recent experiment finds the plateau to be correlated with surface states \cite{dc2}. The main message here is that a steep increase in the resistivity with decreasing temperature does not automatically imply that the system is an insulator, i.e., the chemical potential lies in the gap.

\begin{figure}
\includegraphics[angle=0,width=0.95\columnwidth]{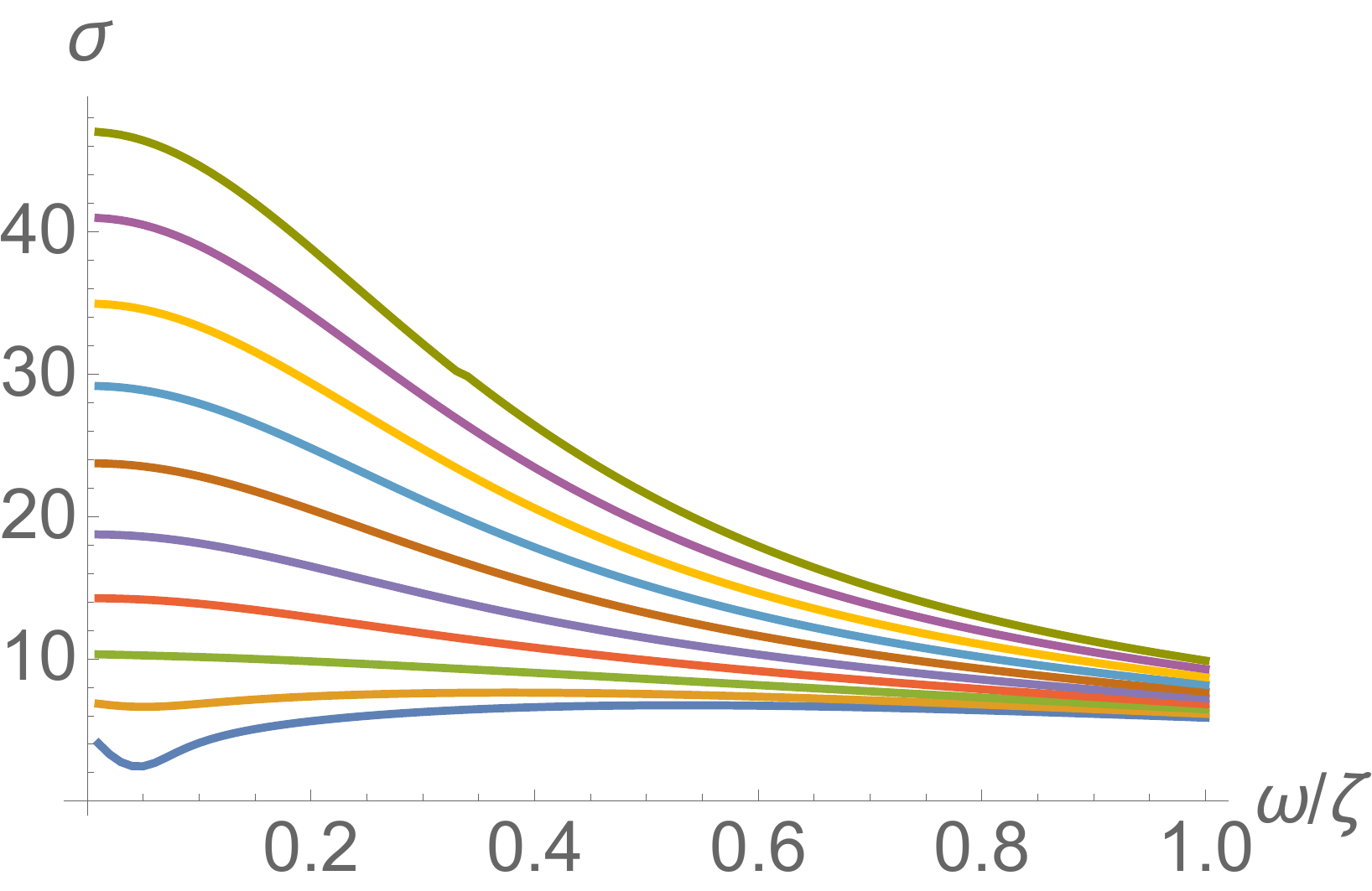}
\caption{Ac conductivity $\sigma$ (arb. units) vs frequency $\omega$ at different temperatures according to Eq.~(\ref{sigmaom1}). Temperature ranges from $0.2\zeta$ to $.01\zeta$ decreasing downwards in steps of $0.021\zeta$. Here, $\zeta\tau=2.5$ with other parameters same as in Fig.~\ref{dctransport}.}
\label{actransport}
\end{figure}

\subsection{Ac conductivity}

The ac conductivity comprises two parts: an interband part and an intraband part. At zero temperature, momentum and energy conservation imply that the interband component of conductivity appears only for frequency $\omega\ge 2\zeta$. Motivated by experiment (see below), I consider the regime $\omega,T<2\zeta$, where only the intraband component contributes. The real part of the ac conductivity is then given by 
\begin{eqnarray}
\sigma(\omega)&=&\sum_{i}e^2\pi\int d\xi \left(\frac{f(\zeta)-f(\zeta+\omega)}{\omega}\right)\nonumber\\
&\times&\frac{1}{(2\pi)^3}\int v_i^2A_i(\mb{k},\xi)A_i(\mb{k},\xi+\omega)d\mb{k},
\label{sigmaom1}
\end{eqnarray}
where $A_i(\ve)=\frac{1/\tau}{(\ve-\ve_{\mb{k}i})^2+(1/2\tau)^2}$ is the spectral function. 
Fig.~\ref{actransport} presents $\sigma$ vs. $\omega$ at different values of $T$. At large $T$, $\sigma$ decreases monotonically with the frequency $\omega$, exhibiting Drude-like response typical of a metal. This is expected since the effect of hybridization is negligible at $T\gg\zeta$. However, at smaller values of $T$, the dependence changes unusually into an increasing function. The effect of disorder is to allow transitions within a width $\sim 1/\tau$ around the Fermi level $\mu$. In a standard metal, the band curvature does not change on the scale of $1/\tau$, therefore the effect is only to produce a reduction in the conductivity. In the present case, however, at low $T$, regions of the band that are less flat become accessible due to the rapid change of band curvature on the scale of $1/\tau$, increasing the velocity of the carriers. This effect now competes with the effect of disorder, leading to an increase in the conductivity with frequency. Such a behavior spans almost the entire range of frequencies, except at extremely low frequencies where an upturn appears--here the increase in carrier velocity is not strong enough to overcome the effect of disorder. This switching from an overall decreasing to an increasing function of frequency is more pronounced when $\mu$ is in the conduction band as opposed to the valence band (not shown in the figure).

An experimental study of the low energy ac conductivity within the hybridization gap of SmB$_6$ was reported in Ref.~\cite{ac1}. There, it was shown that surface states could not account for the origin of the ac conductivity, and raised doubts over whether mid-gap impurity states could explain it either. Remarkably, the behavior shown in Fig.~\ref{actransport} agrees with Fig.~2 in Ref.~\cite{ac1} showing the evolution of $\sigma$ vs. $\omega$ as $T$ is varied: at low $T$, $\sigma$ was found to increase with $\omega$ instead of decreasing. This lends further support to the simple and intuitive picture presented here. 

\begin{figure}
\includegraphics[angle=0,width=0.95\columnwidth]{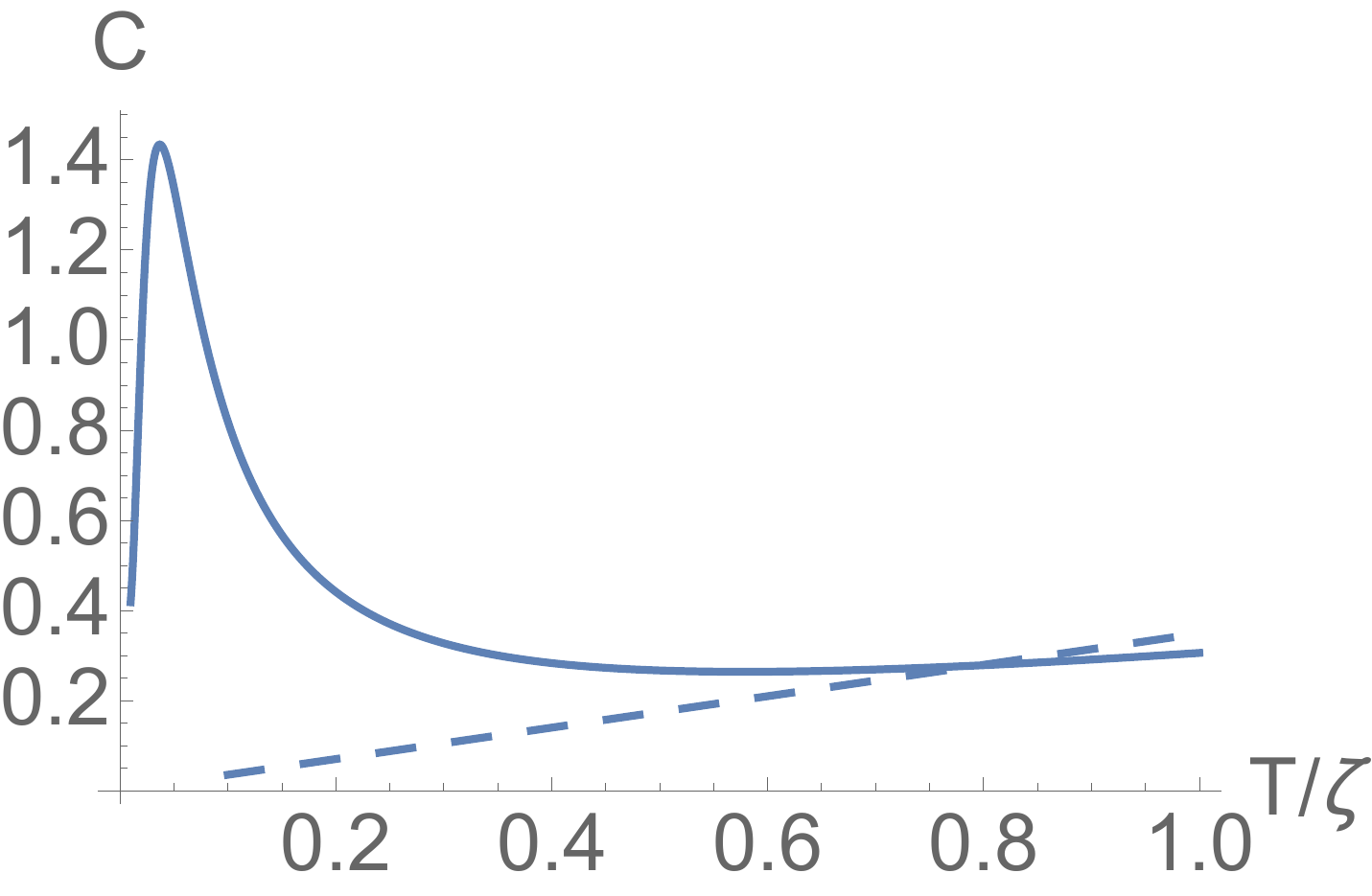}
\caption{Specific heat $C$ (arb. units) vs temperature $T$ according to Eq.~(\ref{speq}). The dashed curve shows the same for the unhybridized parabolic band. Parameters same as in Fig.~\ref{dctransport}.}
\label{specificheat}
\end{figure}

\subsection{Specific heat}

The specific heat is calculated as
\beq
C=\sum_{i}\frac{2}{(2\pi)^3 }\int\ve_i\left(\frac{\partial f_0}{\partial T}\right)d\mb{k}.\label{speq}
\eeq
Fig.~\ref{specificheat} presents the variation of $C$ with $T$. At $T\gg\zeta$ the behavior is similar to that of a standard metal, $C\propto T$, and hybridization has no effect. On the other hand at small temperatures, $T\ll \mu-E_c\ll\zeta$, the system is again metallic, i.e., $C\propto T$, albeit with a slope that is considerably steeper, reflecting the higher density of states. Connecting these two behaviors results in a curve that is nonmonotonic. This is the region where the Sommerfeld expansion valid for conventional metals breaks down since the band curvature changes on a scale comparable with temperature.

It has been known for a long time experimentally \cite{spht1}, and supported by recent experiments as well \cite{spht2}, that SmB$_6$ exhibits a strikingly large value of specific heat which is orders of magnitude larger than conventional metals. How this is possible in spite of being an insulator has escaped a satisfactory explanation. The picture presented here leads naturally to a large value of specific heat at $T<\zeta$, see Fig.~\ref{specificheat} where the specific heat for the present model is compared with that of the unhybridized parabolic electrons. However, it must be noted that the T-dependent curve does not match entirely with what is observed in experiments (although nonmonotonicities are observed in experiments as well). This is not surprising since only the electronic part of the specific heat has been calculated here. Contributions from other degrees of freedom, such as phonons and spin, left out in this calculation, have been shown to be important in fitting experimental curves \cite{spht1,spht2}. Including such contributions will require a microscopic calculation outside the scope of this work. 

\subsection{Quantum oscillations}

In metals, a changing magnetic field causes the Landau levels to cross the Fermi level periodically. This gives rise to oscillations in physical observables, called quantum oscillations. The salient features of these oscillations are as follows: oscillations are periodic in inverse field;  their frequency is proportional to the area of the orbit on the Fermi surface in $k-$space perpendicular to the field; and temperature does not affect the frequency but damps the amplitude in a universal way given by Lifshitz-Kosevich theory \cite{shoenberg}. 

\begin{figure}
\includegraphics[angle=0,width=0.95\columnwidth]{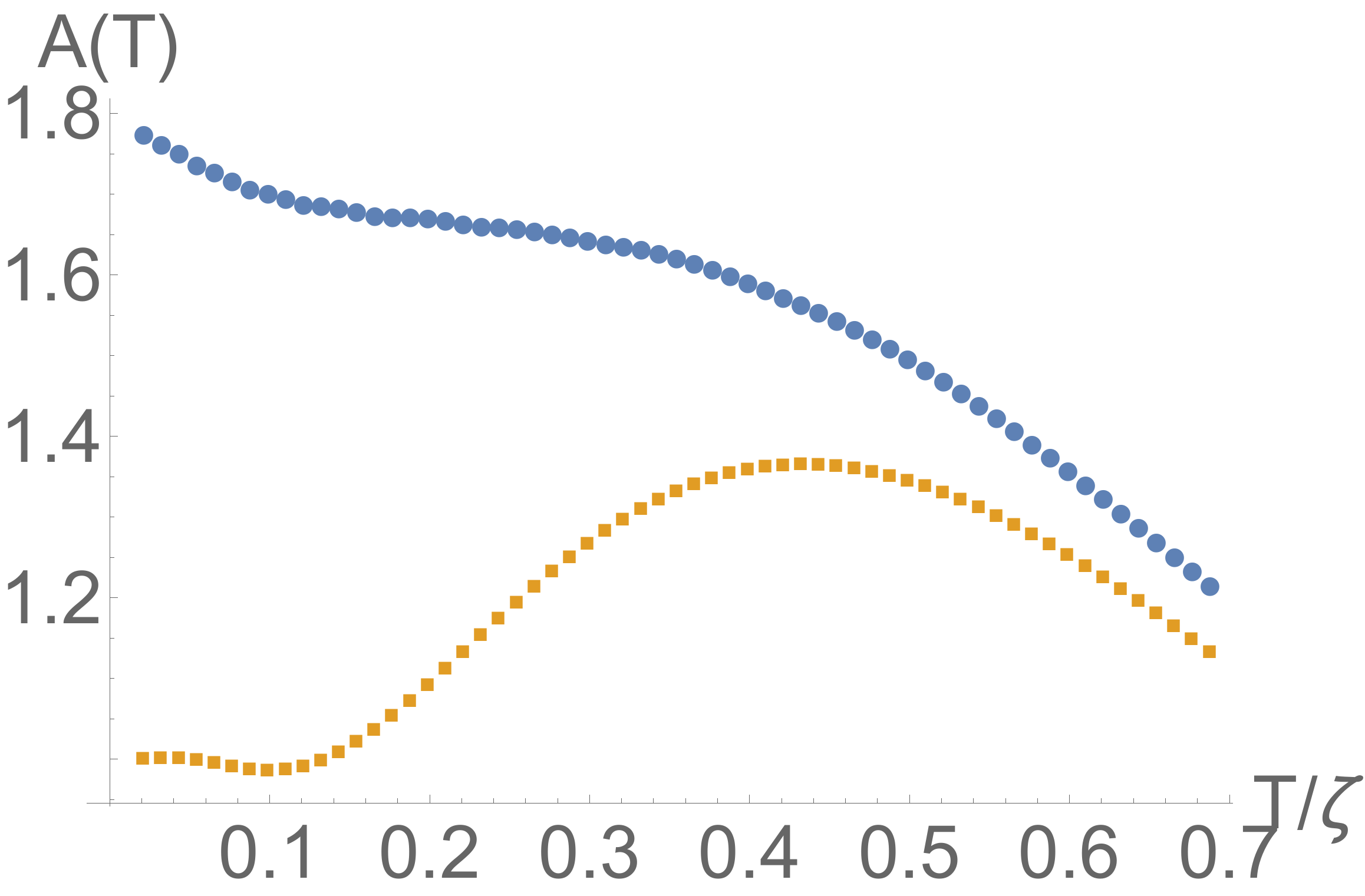}
\caption{Temperature dependence of the amplitude of quantum oscillations for $\zeta/\Delta=0.05$: circles represent $\mu$ in the conduction band but close to the edge ($\mu/\zeta=0.6$) and squares represent $\mu$ in the gap ($\mu=0$). The upturn in the former is clearly visible. All amplitudes are divided by the amplitude at $\mu=0$ and $T=0$. The calculations were done for a lattice model that mimicks the Hamiltonian in Eq.~(\ref{ham}). For details on the computational method, see Ref.~\cite{pal2} (Supplementary materials therein).}
\label{qosct}
\end{figure}

Since the system considered here is a metal, quantum oscillations are expected to appear. However, the rapid change of band curvature, as in other quantities considered before, leads to striking departures from the conventional behavior described above. Oscillations arising in the band structure in Fig.~\ref{band} have been studied extensively in the last few years \cite{knolle1,zhang,pal1,pal2,pal3,knolle3}---I summarize the results pertinent to the present case. 

That this system is unusual is already obvious by noting that even when the chemical potential is in the gap, unlike that in Fig.~\ref{band}---i.e., the system is an insulator---oscillations still appear, contradicting conventional understanding. This was first shown in Ref.~\cite{knolle1}; thereafter, in Ref.~\cite{pal2} it was shown that these unconventional oscillations arise from the sudden change of band slope due to hybridization. This happens at the momentum where the bands were degenerate prior to hybridization; the corresponding energy post  hybridization is $\ve=-\zeta$ (see Fig.~\ref{band}). Thus, oscillations arise from $\ve=-\zeta$ inside the band. When $\mu$ is pushed into one of the bands, the following happens: (i) as long as $\mu$ is inside $[-\zeta,\zeta]$, there are then two sources of oscillations: one from $\ve=-\zeta$, the unconventional one, and one from $\ve=\mu$, the conventional one. However, the two contributions are not of equal strength. Very close to the band edge, the conventional oscillation is weak, being proportional to $1/m$, and the dominant one is still the unconventional one \cite{pal3}. As a result, the oscillations do not have a frequency proportional to the area at $\mu$, as expected in a conventional metal, but rather to the area at $\ve=-\zeta$, which is the same as the area at the intersection point prior to hybridization. The upshot is that the frequency of oscillations will appear as if no hybridization has taken place and oscillations are due to the unhybridized parabolic band \cite{knolle1,pal2,pal3}. (ii) As detailed in Ref.~\cite{knolle1}, instead of following the universal temperature dependence valid for metals given by the Lifshitz-Kosevich formula, the dependence in the insulating case follows a different behavior. When $\mu$ is in the gap, the dependence is nonmonotonic which changes once $\mu$ enters the band. However, as long as $\mu$ is close to the edge, it does not follow the metallic behavior; instead there is a sharp upturn. This is demonstrated in Fig.~\ref{qosct}. (iii) The above two features hold for de Haas van Alphen (dHvA) oscillations (oscillations in magnetization). Shubnikov de Haas (Sdh) oscillations (oscillations in resistivity) can arise only from $\mu$ and not inside the band. At the band edge, therefore, it is expected to be much weaker than dHvA oscillations \cite{pal3}. 

On the experimental side, quantum oscillations in magnetization in SmB$_6$ have been observed, but their origin has not been settled. While Ref.~\cite{qosc1} attributes them to 2D surface states, Ref.~\cite{qosc2} has interpreted them to be of 3D bulk origin. In the latter case, (i) the oscillation frequency has been found to match those of LaB$_6$ which is a conventional metal without any hybridization. (ii) Additionally, a sharp upturn in amplitude at low temperatures is observed, deviating from the conventional metallic behavior. (iii) And, unlike magnetization, no oscillations have been observed in the resistivity. These three features are consistent with the corresponding ones listed in the preceding paragraph.

\section{Concluding remarks}

A simultaneous presence of metal- and insulator-like properties in SmB$_6$ has spurred a fierce ongoing debate. Based on paradoxical experimental observations, competing theories and interpretations have appeared. Nevertheless, most of these theories are inspired by a common line of thinking: the gap is exotic with nontrivial excitations. This paper provides an alternative picture that is simpler and intuitive. Until recently, the conventional wisdom on SmB$_6$ has been that the latter is a renormalized band insulator---`renormalized silicon'. The theory presented here is more along this traditional approach. It emphasizes that, at least as far as the anomalies in the four quantities considered here are concerned (and possibly other ones), they could be explained by a simpler mechanism within this traditional approach. Note, however, the theory does not contest recent predictions of topological surface states or the importance of interactions, which could manifest in other ways. 

Additionally, from a pedagogical perspective, the model considered presents an instructive example where textbook results on metals and insulators break down due to certain features in the band.  

It is hoped that the alternative viewpoint espoused above will inspire further work. A much more quantitative theory with a realistic band diagram of SmB$_6$ is necessary to make quantitative comparison to experimental data. For such a comparison, a complete set of experiments on the calculated quantities needs to be done on the same (batch of) samples. And, finally, a microscopic theory justifying the phenomenological calculations above is required.

\begin{acknowledgments}
I would like to thank F. Pi{\'e}chon, M. Rozenberg, M. Civelli, and P. Coleman for valuable discussions at different stages which led to this work, and LabEx PALM Investissement d'Avenir (ANR-10-LABX-0039-PALM) for financial support during the intial stages of this work.
\end{acknowledgments}

\end{document}